\documentclass[fleqn,10pt]{olplainarticle}

\usepackage{url}
\usepackage{graphicx} 
\graphicspath{{figs/}}
\usepackage{float}
\usepackage{algorithm}  
\usepackage{algpseudocode}
\usepackage{xcolor}
\usepackage{setspace}
\usepackage{physics}
\usepackage{subcaption}
\usepackage[normalem]{ulem}

\author[Wouter Deleersnyder]
{Wouter Deleersnyder$^{1,2}$,  David Dudal$^{1,3}$, Thomas Hermans$^{2}$\\
	$^{1}$KU Leuven Campus Kortrijk - KULAK, Department of Physics, Etienne Sabbelaan 53, 8500 Kortrijk, Belgium.\\  E-mail: {wouter.deleersnyder@kuleuven.be} \\
	$^{2}$Ghent University, Department of Geology, Krijgslaan 281 - S8, 9000 Gent, Belgium \\
	$^{3}$Ghent University, Department of Physics and Astronomy, Ghent, Krijgslaan 281 - S9, 9000 Gent, Belgium \\
}

\title{A multidimensional AI-trained correction to the 1D approximate model for Airborne TDEM sensing}


\begin{abstract}
	The computational resources required to solve the full 3D inversion of time-domain electromagnetic data are immense.  To overcome the time-consuming 3D simulations, we construct a surrogate model, more precisely, a data-driven statistical model to replace the 3D simulations. It is trained on 3D data and predicts the approximate output much faster. We construct a surrogate model that predicts the discrepancy between a 1D subsurface model and a deviation of the 1D assumption. The latter response is fastly computable with a semi-analytical 1D forward model. We exemplify the approach on a two-layered case. The results are encouraging even with few training samples. Given the computational cost related to the 3D simulations, there are limitations in the number of training samples that can be generated. In addition, certain applications require a wide range of parameters to be sampled, such as the electrical conductivity parameters in a saltwater intrusion case. The challenge of this work is achieving the best possible accuracy with only a few thousand samples. We propose to view the performance in terms of learning gain, representing the gain from the surrogate model whilst still acknowledging a residual discrepancy. Our works open new avenues for effectively simulating 3D TDEM data.
\end{abstract}

\begin{document}
\flushbottom

\maketitle
\textbf{\color{red} Published in \textit{Computers \& Geosciences}:\\
Deleersnyder, W., Dudal, D., \& Hermans, T. (2024). A multidimensional AI-trained correction to the 1D approximate model for Airborne TDEM sensing. \textit{Computers \& Geosciences}, 188, 105602.}
\thispagestyle{empty}

\section{Introduction}
\label{sec:introduction}
Airborne Electromagnetic ({AEM}) mapping is an increasingly used method to image near-surface geological features over large areas via the bulk electrical resistivity. Common applications are, for example, in mineral exploration \citep{macnae2007developments}, contamination \citep{pfaffhuber2017delineating}, hydrogeological mapping \citep{mikucki2015deep}, and saltwater intrusion \citep{siemon2019automatic, deleersnyder2023flexible}. Part of the AEM systems collects time-domain electromagnetic ({TDEM}) data, but those systems require significantly more computer power to run 3D modelling simulations, compared to frequency-domain electromagnetic ({FDEM}) data. 3D FDEM forward modelling is becoming applicable for smaller areas \citep{ansari2017gauged, cockett2015simpeg,haber2004inversion, werthmuller2019emg3d}, while 3D TDEM forward modelling remains impractical \citep{borner2015three, haber2007inversion}. For example, \citet{engebretsen2022accelerated} could carry out a 2D TDEM inversion with 418 soundings, which took 30 hours, 97 GB of memory for only 16 iterations. \citet{auken2017review} argues that the computational demand for a full 3D AEM survey is still too large so that only few companies are capable of realizing such inversions.\\

By consequence, quasi-2D/3D inversion methods, using a much faster and approximate (1D) forward model,  methods are standard for large-scale mappings. The 1D forward model assumes cylindrical symmetry and is thus limited to horizontally layered conductivity models. Using that forward model in a quasi-2D/3D method may yield erroneous interpretations and the results may better not be interpreted quantitatively. \citet{deleersnyder2022novel} provide an appraisal tool which verifies whether an erroneous interpretation may occur as a result of the approximate 1D forward model, and indicates areas that should be reinterpreted with a full 3D forward model.\\

To overcome the time-consuming simulations, a common data-driven approach is to construct surrogate models. It is a statistical model that can approximate the simulation output. A single evaluation of the trained statistical model is much faster than a simulation. The statistical relation is built from pairs of full simulation data and its inputs. Surrogate modelling is thus a case of supervised machine learning. Surrogate modelling and machine learning may replace 3D forward modelling on a mesh during a 3D inversion, sometimes coined Fast Forward Modelling (FFM).\\

The existing literature of using Machine Learning ({ML}) and 1D forward modelling is already significant. \citet{bording2021machine} use a conventional Neural Network ({NN}) for ground-based TDEM data and claims a factor 13 speed up compared to the accurate 1D forward models with 98\% accuracy within the 3\% relative error limit. Other examples of 1D surrogate models are found in surface wave dispersion \citep{hou2019learn}, seismic and marine controlled-source EM data \citep{puzyrev2021inversion} and seismic full-waveform inversion \citep{wu2018inversionnet}. For AEM modelling, \citet{asif2022integrating} train a NN for 1D forward modelling, including the flight altitude as additional parameter, which adds complexity to the learning problem. Additionally, two separate NNs were trained, one to predict the forward response and one for the partial derivatives. A speed-up with a factor 50 is claimed. \citet{wu2023deep} have trained an alternative NN with a long short-term memory (LSTM) architecture with similar `functionalities' and claim a factor 2700 speed-up. The step towards 3D modelling has already been made for gravity data \citep{zhang2021deep} or frequency domain magnetotellurics \citep{conway2019inverting, peng2023rapid}, which suffer less from the computational burden to generate training data. \\

Other work shows that NNs can be used for a direct inversion \citep{noh2020imaging, feng2020resistivity, bai2020quasi, li2020bp, puzyrev2021inversion, laloy2021approaching}. This avoids the potentially time-consuming optimization step of the inverse problem, but one loses control over the specific tunable aspects of the inversion process. The trained NN needs to handle the non-uniqueness of the geophysical problem, as well as the noise. By training a separate forward/surrogate model, one retains control over the inversion: The type of regularization and the regularization parameter(s) can be easily modified. This allows to use the trained surrogate forward model in flexible regularization schemes where, for example, the sharpness can be tuned \citep{deleersnyder2023flexible, klose2022laterally}.\\

While much work has been done in exploring the surrogate modelling and Machine Learning for 1D forward problems in TDEM problems, the step towards multidimensional forward modelling, to our knowledge, has not yet been made. While the literature above has used up to a million samples in the training dataset, this is not feasible for 3D TDEM simulations. The challenge of this work is achieving the best possible accuracy with only a few thousand (5090) samples. We, therefore, limit the parameter space to subsurface models with two layers.
Rather than predicting the 3D data directly, we predict the relative error between the 3D and 1D data \citep{deleersnyder2023machine}. The computation of the 1D approximate forward model is fast and explains most of the variability in the geophysical data. The discrepancy, originating from the multidimensional nature of the subsurface, is predicted with machine learning tools. We focus on a minimal, well-defined context, namely a central-loop airborne system flying at 40 m altitude, covering an area with high conductivity contrasts, such as when imaging a fresh saltwater interface. \\

We have opted for more classical ML methods which are well-established in geosciences and are more insightful, rather than using (advanced) neural networks that are often data hungry for the learning step. We will use Gaussian Process Regression, typically used in forecasting \citep{ferkous2021wavelet, satija2015direct}, kriging \citep{goncalves2022learning, ching2023data} or in geohazard mitigation applications \citep{dikshit2021pathways, gao2022landslide}. We work with the common Principal Component Analysis (PCA) as dimensionality reduction technique, which finds many applications \citep{grana2019uncertainty, scheidt2018quantifying,michel20201d}. Functional Component Analysis, e.g., used in \citet{satija2015direct}, is less common in geosciences \citep{suhaila2023research} as a dimensionality reduction technique. Demonstrating the performance of a surrogate model for a simplified case without advanced ML leaves room to use more advanced ML in more extensive contexts (with more parameters).\\

We first discuss the data generation in Section \ref{sec:sub:data} in terms of low and high fidelity data, the 1D forward model and  3D simulations respectively. In Section \ref{sec:subsub:datatransformations}, we discuss the data transformations used in this work. Inspired by the characteristics of the transformed data (discussed in Section \ref{sec:sub:insights}) and preliminary tests, a Gaussian Process Regression ({GPR}) is picked as surrogate model (Section \ref{sec:sub:surrogate}). In Section \ref{sec:sub:performance}, the performance of the trained surrogate models is described, suggesting successful routes for future work. 

\section{Methods}
\label{sec:methods}
\subsection{Data generation}
\label{sec:sub:data}
The objective of this work is to construct an ML-based correction model on an incomplete, but fast evaluable low-fidelity forward model. For TDEM data we use a 1D semi-analytical model \citep{hunziker2015electromagnetic}, which assumes horizontal layers, to generate Low Fidelity (LF) data. A Python, open-source implementation is used  \citep{werthmueller2017open} .  The correction model is trained using High Fidelity ({HF}) data, which must be generated by a forward model that can accommodate multidimensional structures. The only viable method is through numerical simulations, described in Section \ref{sec:subsub:highfidelity}. Data transformations, before training, are described in Section \ref{sec:subsub:datatransformations}.\\

The surrogate correction model predicts the target with the following listed features: 
\begin{itemize}
	\item The depth $d$ (in meter) of the interface between the two layers, $d \in [10, 50]$,
	\item the angle $\theta$ (in $^\circ$) of the dipping layer (0° being purely horizontal), $\theta \in [-75, 75]$ and,
	\item the Electrical Conductivity of the first ($EC1$) and second ($EC2$) layer (in~Sm$^{-1}$), $EC1, EC2 \in [0, 0.75]$..
\end{itemize}
To test the generalizability of this approach, the prior (or the training data) is not specifically adapted to a specific application. The data points are sampled uniformly within the feature space within the specified boundaries.

\subsubsection{Finite Volume Method simulations to generate High Fidelity data}
\label{sec:subsub:highfidelity}
The finite volume method ({FVM}) is a numerical discretization technique for representing and evaluating partial differential equations by conversion to algebraic equations. The FVM evaluates exact expressions for the average value of the quantities over some (finite) volume and uses this discretization to construct an approximate solution. In this work, we resort to the SimPEG package \citep{cockett2015simpeg, heagy2017framework}, which is an open-source project in Python which implements the Maxwell's equations and the most common geophysical electromagnetic sources, and the parallel, direct PARADISO-solver \citep{schenk2004solving}.\\

\begin{figure}
	\centering
	\includegraphics[width=0.7\linewidth]{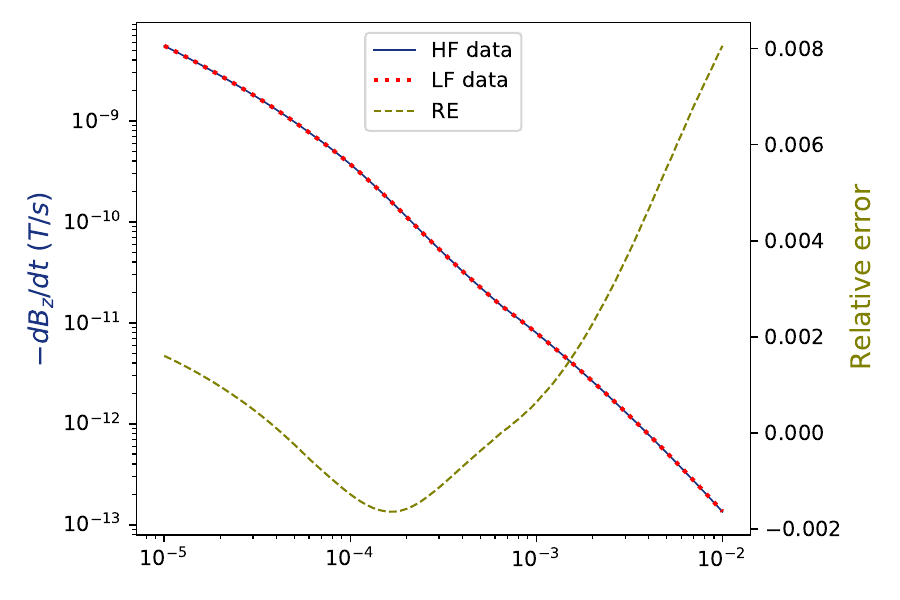}
	\caption[Comparison between high fidelity and low fidelity data.]{Illustrative comparison of Low Fidelity ($LF$) and High Fidelity ($HF$) data with horizontal layers for 0.1 and 0.4 Sm$^{-1}$ and an interface at 50 m depth illustrate the accuracy of the $HF$ mesh. The relative discrepancy between the $HF$ and $LF$ data is also shown; it indicates larger errors at later times.}
	\label{fig:comparisonlfvshf}
\end{figure}

The High Fidelity mesh in this work is a 3D regular rectangular mesh with 288~000 cells. The core ranges from $-50$ to $50$ m in the lateral orientations and from $-140$ to $60$ m in the vertical orientation with cell sizes of $5$-by-$5$ m. Outside the core region, cells are progressively increased with a rate of 20\% up to a cell size of $200$ m. The total mesh extent is 2,35 km in the lateral orientation and 2,80 km in the vertical orientation. The accuracy of the $HF$ data is illustrated on a two horizontally layered case with 0.1 and 0.4 Sm$^{-1}$ and an interface at 50 m depth. The $HF$ and $LF$ data are shown in Figure \ref{fig:comparisonlfvshf}. It is unavoidable to have a small discrepancy. In this case the relative discrepancy averages 0.19\% per time gate and a maximum of 0.8\% at late time gates; see Figure \ref{fig:comparisonlfvshf}.

\subsubsection{Dataset and transformations}
\label{sec:subsub:datatransformations}
The training and test dataset consists of 5090 and 566 samples, respectively. The target variable $\frac{\partial \vb{B}}{\partial t}$ is sampled logarithmically from $10^{-5}$ to $10^{-2}$ seconds with 100 time steps. The target output is thus a vector of 100 data points, calling for a dimensionality reduction, which is described in Section \ref{sec:subsub:dimensionality}. Before the dimensionality reduction, the magnetic data is transformed to a relative error $RE$ between the $LF$ and $HF$ data.
\begin{equation}
	RE = \frac{HF - LF}{LF}
\end{equation}
By predicting the error, solely the residual multidimensional variability has to be learned. Transforming the target to a relative error, rather than an absolute error, ensures that the target outputs are within the same range, as the original magnetic data (and consequently the absolute noise) spans multiple orders of magnitude.

The sensitivity of the forward response does not vary linearly with changes in the conductivity. \citet{bording2021machine} suggests the following normalization for a better optimization:
\begin{equation}
	\tilde{EC} = a + \frac{(b-a)\left(\log_{10}EC - \log_{10}(\min EC\right)}{\log_{10}\max(EC) - \log_{10}\min(EC)},
\end{equation}
where $\tilde{EC}$ is the normalized electrical conductivity. The minimum and maximum correspond to the minimum and maximum electrical conductivity in the training dataset. The values are transformed within the interval $[a,b]$, which is $[-1,1]$ in our case. 

Before training and after the dimensionality reduction of the target, all features and target outputs are standardized (by substracting the mean and dividing by the standard deviation). This is a common requirement for many machine learning estimators.
\subsection{Surrogate modelling}
\label{sec:sub:surrogate}
\subsubsection{Dimensionality Reduction}
\label{sec:subsub:dimensionality}
The target consists of 100 variables. However, the smoothness of the underlying physics implies that neighbouring variables are correlated, i.e., the $HF$ data varies smoothly throughout time, which calls for a dimensionality reduction technique. Dimensionality reduction techniques transform high-dimensional data into a lower dimensional space, retaining the most descriptive features, depending on the technique considered. Dimensionality reduction is often used in machine learning, as it combats the computational cost and is able to control overfitting \citep{lopez2021deep}.

In this work, the target variables are consecutively transformed to the functional and principal component space. This is called \textit{Functional Component Analysis} (FCA) \citep{silverman1996smoothed,satija2015direct}.  First, the target is represented in a functional data space via Functional Data Analysis (FDA) \citep{ramsay2006functional}, which represents the original data as a linear combination of differentiable mathematical functions:
\begin{equation}
	\label{eq:FDA}
	y(t) \simeq \sum_{k}^{K} c_k \phi_k(t),
\end{equation}
where $\phi_k(t)$ is a set of basis functions. The true, underlying basis of our target is unknown. Therefore, we view each functional datum (each sample), using a non-parametric representation, meaning that each value of the 100 target variables are stored in a finite grid of points. The underlying physics is smooth and thus the functional data are smooth and one may use interpolation to evaluate between the grid. With FDA, we represent that data in a parametric fashion, using the linear combination of Eq. \eqref{eq:FDA}. Common systems rely on the Fourier basis, monomial basis, finite element basis etc. We work with the B-spline basis. It can represent smooth functions in a sparse fashion and it is computationally efficient.

We can now reduce the dimensionality, by representing the non-parametric (original) data in a parametric fashion, i.e. coefficients of the corresponding spline basis, where the number of coefficients $K < 100$ can still relatively accurately represent the original data.

On those functional data coefficients, Principal Component Analysis (PCA) is performed, providing a representation in terms of the principal vector basis. It consists of a linear mapping which maximizes the variance of the data in that principal vector basis or principal components. Those principal components are linearly uncorrelated and are ordered by decreasing variance. By excluding the principal components which explain little variance, the dimensionality can be further reduced. We denote the number of retained components with $L$. 

Note that FCA is different than the method coined Functional Principal Component Analysis (FPCA), which retains modes in functional variation of longitudinal data \citep{hall2006properties}.

\subsubsection{Gaussian Process Regression}
\label{sec:subsub:GPR}

In Gaussian Process Regression (GPR), the concept of a Gaussian Process (GP) is used for making predictions. \citet{williams2006gaussian} defines a GP as a collection of random  variables, any finite number of which have a joint Gaussian distribution and is fully specified by second order statistics: a mean and a covariance function. In a GPR context, the latter is referred to as the \emph{kernel}. The kernel encodes the type of structure that is found in samples drawn from the GP prior by defining the “similarity” of two datapoints combined with the assumption that similar datapoints should have similar target values. When making a prediction, we evaluate the samples at a given point, but we must first incorporate the knowledge from the training data \citep[p.~77]{scheidt2018quantifying}.

Our proposed model is a GPR and relies on the Radial Basis Function (RBF) in combination with a White Noise kernel. The RBF is a squared-exponential kernel
\begin{equation}
	\label{eq:RBF}
	k_{\text{RBF}}(x_i, x_j) = \exp\left(-\frac{d(x_i, x_j)^2}{2\ell^2}\right),
\end{equation}
where $\ell$ is the length-scale of the kernel and $d(\cdot,\cdot)$ the Euclidean distance. It is a default kernel with appealing properties (it is stationary, i.e., it does not directly depend on the value of $x$, but the distance between two points, and it has infinitely many derivatives) \citep[p.~9]{duvenaud2014automatic}. The RBF generates smooth samples in the GP prior and is generally a good choice if no discontinuities are expected, as in our case. Increasing the length scale leads to more slowly varying GPs. 

The White Noise kernel is included,
\begin{equation}
	k_{\text{WN}}(x_i, x_j) = \left\{
	\begin{array}{ll}
		\sigma^2 & \mbox{if } x_i = x_j  \\
		0 & \mbox{else}
	\end{array}
	\right.
\end{equation}
meaning that the GPR will explicitly learn the variance $\sigma^2$ of the noise, rather than to solely rely on the Tikhonov regularization (smoothing) which is assumed on the training point's covariance matrix, as the covariance matrix may be ill-conditioned, controlled by parameter $\alpha$ within the GPR, it works similarly as the White Noise Kernel, but the variance is not a free parameter). The main origin of the noise in the dataset is due to the discretization of the subsurface model for the $HF$ fidelity data. For example, when the interface at depth $d$ does not coincide with the edges of a cell, the depth is rounded to coincide with the edge of that cell, causing an apparent random noise. We are not interested in the noise and want to prevent overfitting, such that the fitted length scales of the RBF kernel follow the slowly varying, general trend, rather than the short-scale variations due to the noise. The latter would limit the interpolating power of the GPR.

We train a separate GPR for each functional component, without using a multiple-output GPR. The latter could exploit output correlations, however, we train in the functional component space, where PCA was used to create uncorrelated functional components. In Section \ref{sec:discussion}, we discuss the simplicity of using single-output GPRs over one multiple-output GPR which has the advantage of considering correlation between responses. This means that we will optimize for each functional component separately. The length scales and the noise-level will potentially be different for each component.

\section{Results}
\label{sec:results}
\subsection{Insights from transformed data}
\label{sec:sub:insights}
\begin{figure}
	\centering
	\begin{subfigure}[b]{.49\textwidth}
			\centering
			\includegraphics[width=\textwidth]{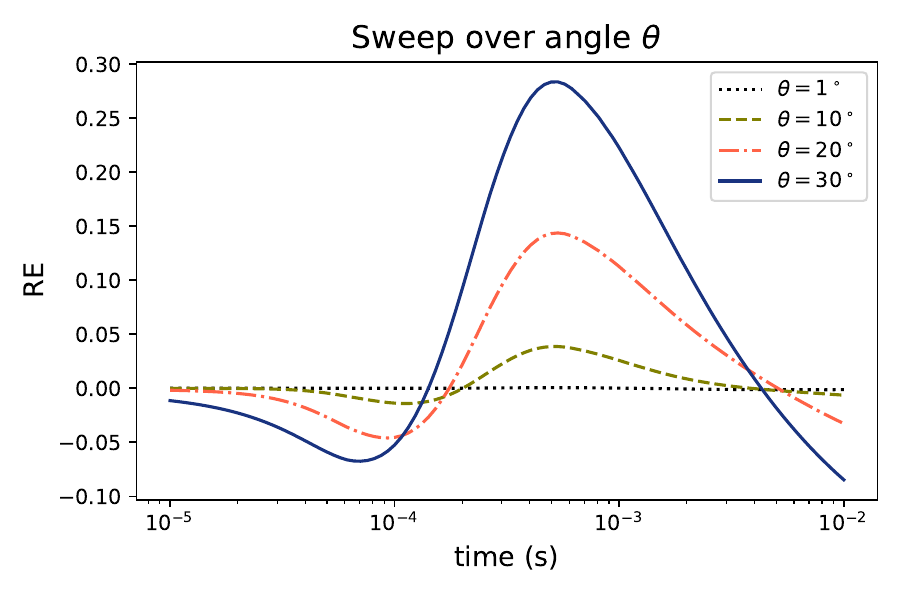}
			\caption{Sweep over angle $\theta$. The Relative Error (RE) increases with increasing angle $\theta$.}
			\label{fig:insightsa}
	\end{subfigure}
	\begin{subfigure}[b]{.49\textwidth}
		\centering
		\includegraphics[width=\textwidth]{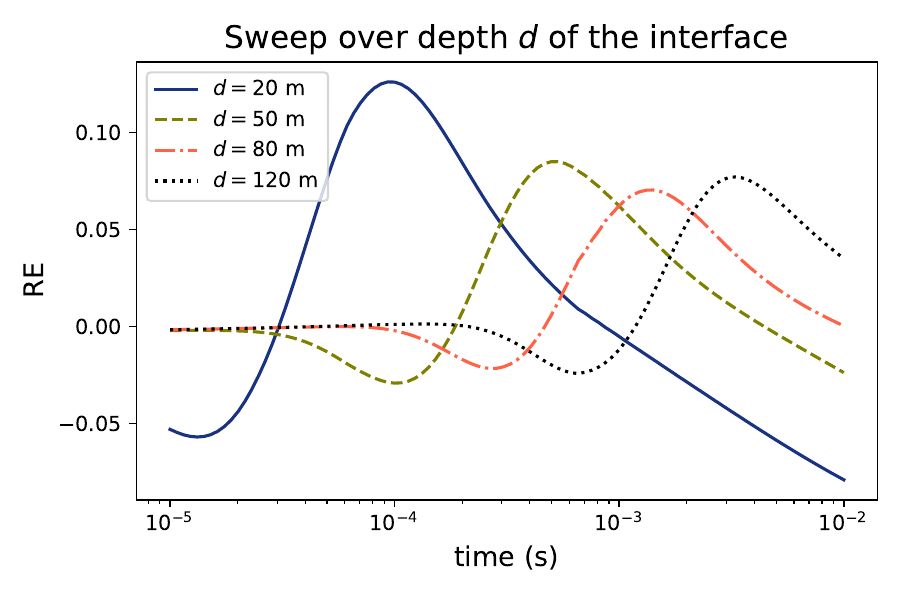}
		\caption{Sweep over depth $d$ of the interface. The peak of the max Relative Error (RE) shifts to later times for larger depths $d$.}
		\label{fig:insightsb}
	\end{subfigure}
	\begin{subfigure}[b]{.49\textwidth}
		\centering
		\includegraphics[width=\textwidth]{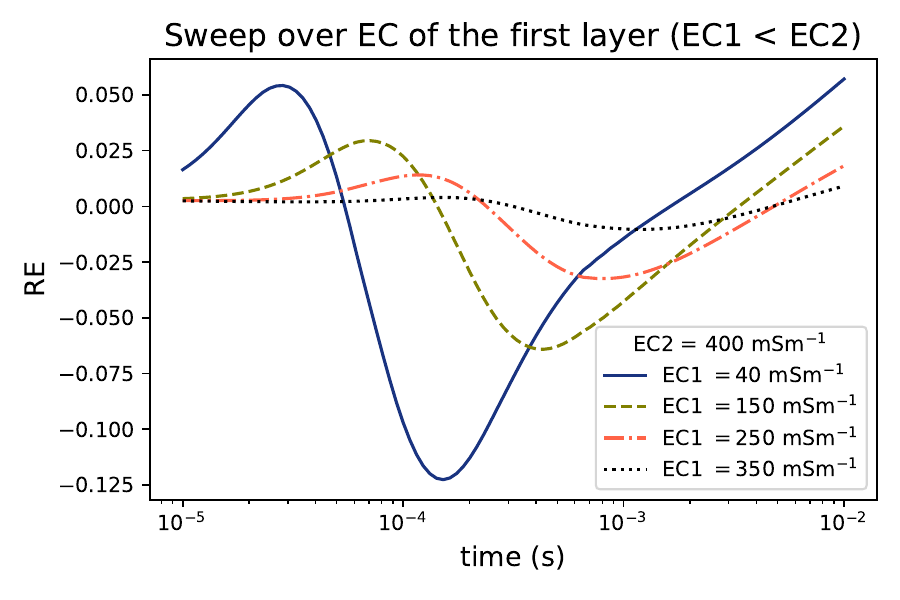}
		\caption{Sweep over the electrical conductivity of the first layer, in the case where the conductivity of the first layer is lower than the conductivity of the second layer.}
		\label{fig:insightsc}
	\end{subfigure}
	\begin{subfigure}[b]{.49\textwidth}
		\centering
		\includegraphics[width=\textwidth]{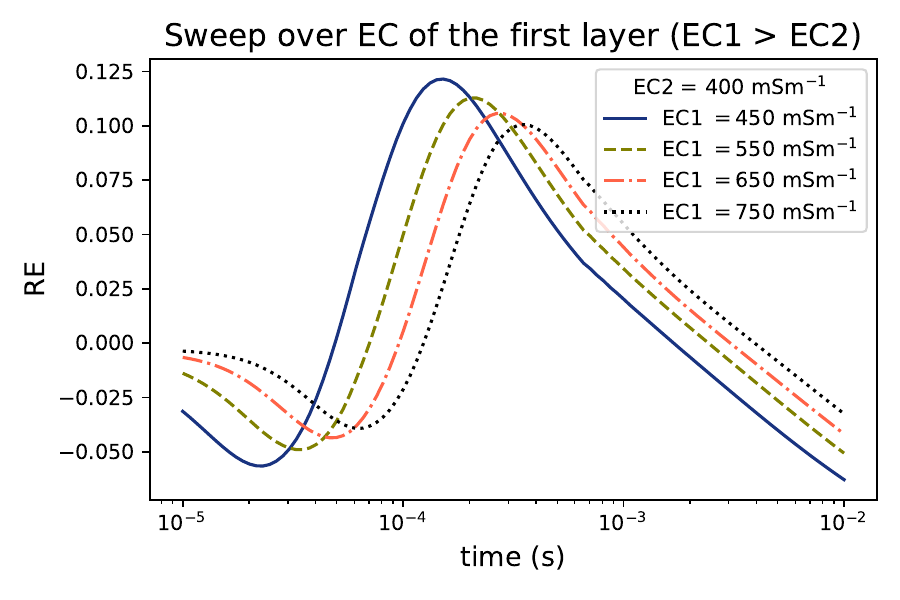}
		\caption{Sweep over the electrical conductivity of the first layer, in the case where the conductivity of the first layer is layer than the conductivity of the second layer.}
		\label{fig:insightsd}
	\end{subfigure}
		\begin{subfigure}[b]{.49\textwidth}
		\centering
		\includegraphics[width=\textwidth]{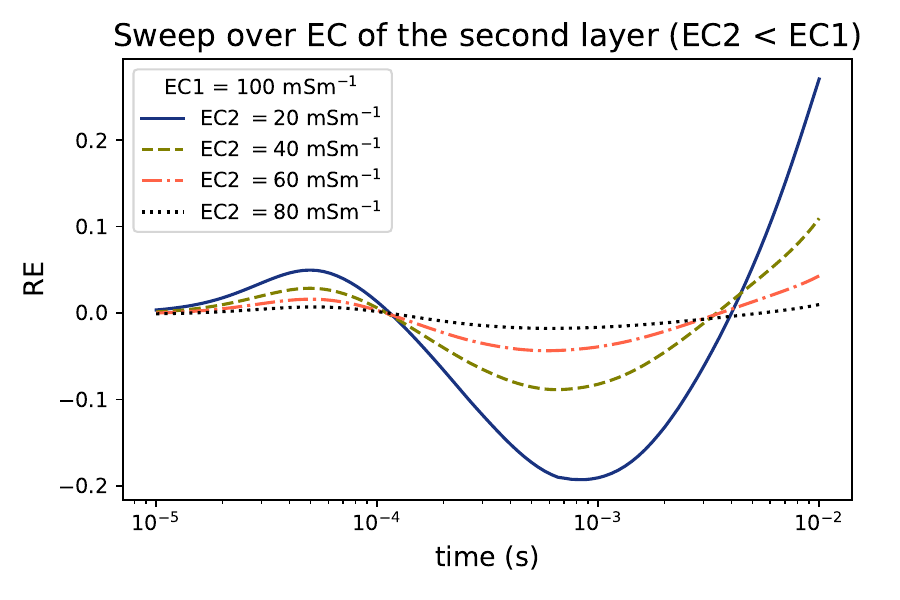}
		\caption{Sweep over the electrical conductivity of the second layer, in the case where the conductivity of the second layer is lower than the conductivity of the first layer.}
		\label{fig:insightse}
	\end{subfigure}
	\begin{subfigure}[b]{.49\textwidth}
		\centering
		\includegraphics[width=\textwidth]{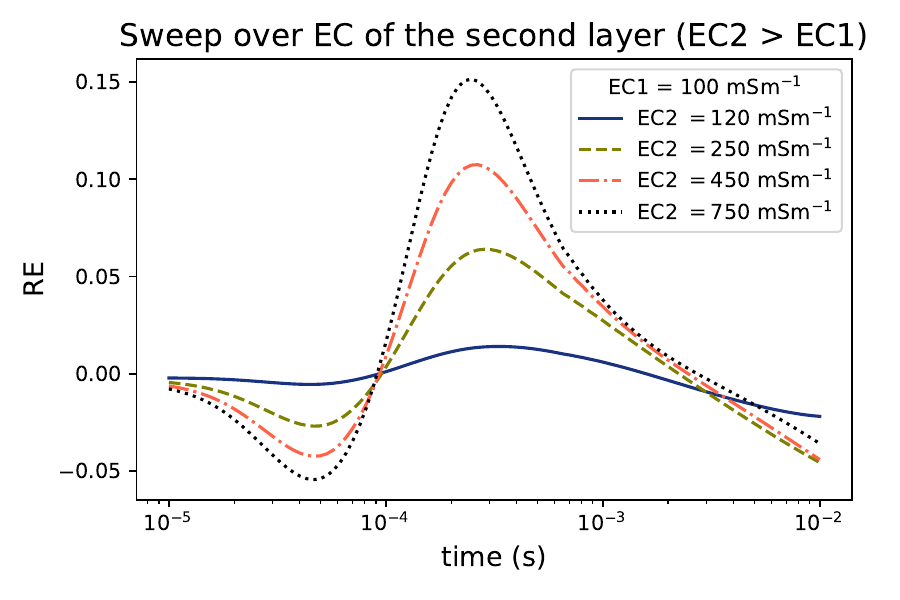}
		\caption{Sweep over the electrical conductivity of the second layer, in the case where the conductivity of the second layer is larger than the conductivity of the first layer.}
		\label{fig:insightsf}
	\end{subfigure}
	\caption[Relative error curves in a parameter study.]{Relative error (RE) curves in a parameter study. In each plot,  only one parameter is varied at a time. The fixed parameters were $\theta = 15^\circ$,  $d = 50$ m, EC1 = 100 mSm$^{-1}$, EC2 = 400 mSm$^{-1}$. }
	\label{fig:insights} 
\end{figure}

We can easily apply the interpretation of the diffusing currents to the obtained relative error ({RE}) curves in Figure \ref{fig:insights}. The peaks in the RE can be explained by the bulk of the current distribution reaching the interface, where the subsurface deviates from the horizontally layered models. If we perform a sweep over the angle, we find that larger angles yield larger maxima in the RE. A sweep over the depth shows, not unexpectedly, that a deeper interface yields a maximum at later times, because the diffusing currents travel longer to reach the discrepancy at the interface. The EC of the first layer has a large impact on the `arrival time' of these diffusing currents. We know that electrical currents diffuse faster in lower conductive media, which is confirmed in the RE curves. The EC in the second layer does not impact the time of the peaks in RE, but the figure nicely illustrates that the contrasts between two electrical conductivities are more significant to explain the value of the RE peaks, rather than the actual electrical conductivities itself.

We need to check to what extent the target can represent the original relative error after dimensionality reduction. With $L<100$, the number of functional components will no longer be able to represent all relative errors $RE$ accurately. For this analysis, the dimensionality reduction was trained on the training dataset and its performance, expressed in Mean Absolute Error ({MAE}), tested on the test dataset. The MAE is computed between the original data ($RE$) and the back transform of the reduced data $\tilde{RE}$,
\begin{equation}
	MAE ={\frac {1}{n}\sum _{i=1}^{n}\left|RE_i-\tilde{RE}_i \right|},
\end{equation}
where $n$ is the number of samples in the training dataset. The minimum MAE in this analysis will be the infimum for the surrogate model performance.

\begin{figure}
	\centering
	\includegraphics[width=0.65\textwidth]{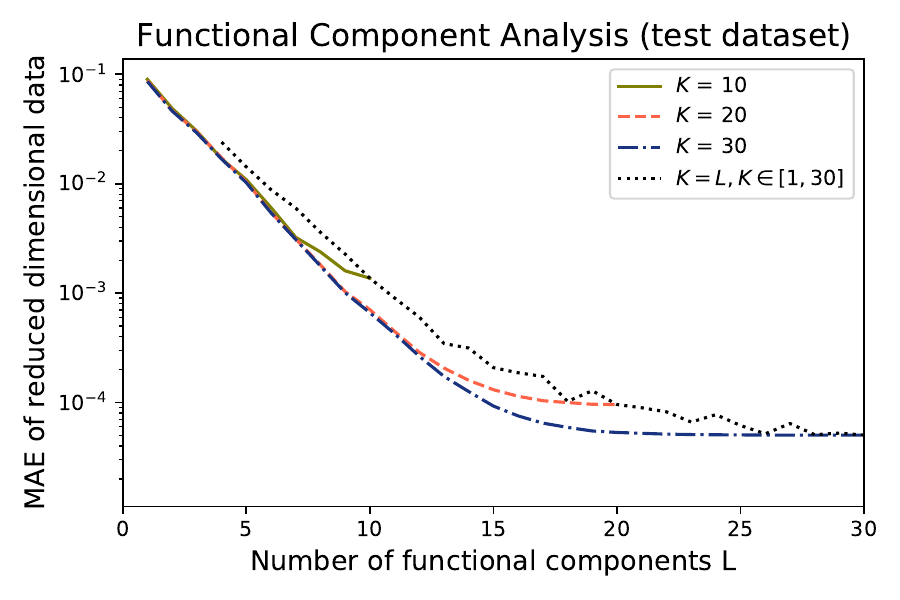}
	\caption[Minimum Mean Absolute Error due to dimensionality reduction.]{Mean Absolute Error (MAE) introduced by working on a reduced dimensional space (rather than inaccurate prediction of the estimator), for different retained functional components $L$ of the $K$ B-spline functions. MAE for $L>30$  is approximately constant and therefore not shown.}
	\label{fig:dimensionality_reduction} 
\end{figure}

With the FDA, the dimensionality is reduced to $K = 10, 20$ and 30 dimensions, meaning that we use $K$ B-spline basis functions to represent our original RE. Then, $L$ functional components are retained after PCA on the coefficients of the FDA. The MAE is shown in Figure \ref{fig:dimensionality_reduction} for each $L \leq K$. Also, the MAE for all $K$ in $[1,30]$, while retaining \textit{all} functional components, is shown (black dotted line). This MAE is identical to the MAE as if you would not perform PCA on the FDA coefficients.

From Figure \ref{fig:dimensionality_reduction}, we observe that an increase in the number of B-spline functions $K$ can better represent the original relative error RE, as expected. However, this effect becomes insignificant for $K > 20$. Note that the minimum number of B-splines is 4. This does not hold for PCA. Secondly, we learn that many functional components are required to represent the original data in a reduced dimensional space accurately. If $K$ is low, the results suggests to retain almost all $L$ functional components. For larger $K$, one can gain quite a bit of dimensionality reduction by using PCA. For example, for $K = 30$, one can easily further halve the dimensions without much loss of accuracy.

We do not decide on the optimal number of B-spline basis functions $K$ and the number of functional components $L$ using this analysis. It will serve as a guide to reduce the optimization for these parameters during the training of the GPR. The MAE of the \emph{predicted} relative error in our original space will be the only criterion to determine $K$ and $L$.

\subsection{Performance}
\label{sec:sub:performance}

\begin{figure}
	\begin{subfigure}[b]{.49\textwidth}
		\centering
		\includegraphics[width=\textwidth]{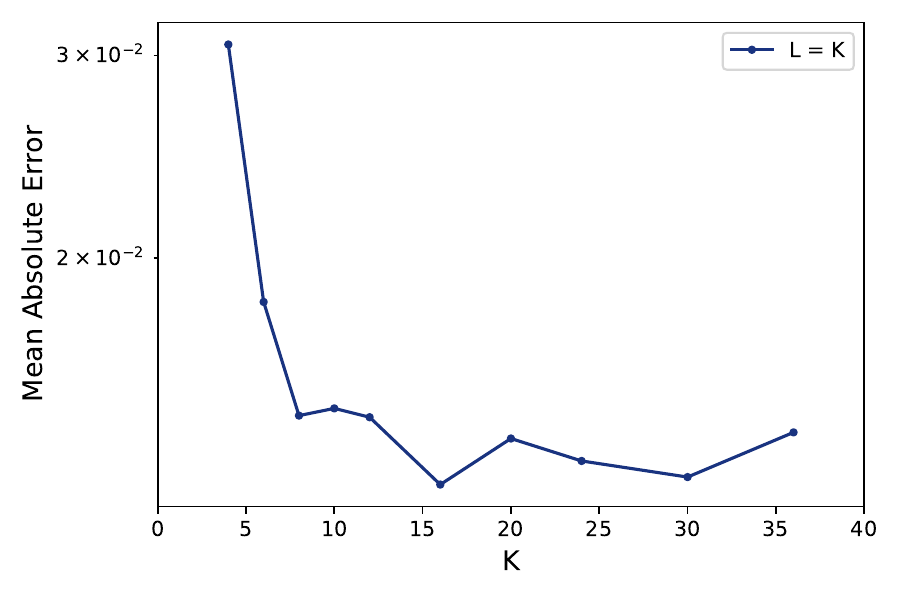}
		\caption{The accuracy for varying the number of functional basis functions $K$ and retaining all functional components ($K=L$).}
		\label{fig:MAE_in_terms_ofa}
	\end{subfigure}
	\begin{subfigure}[b]{.49\textwidth}
		\centering
		\includegraphics[width=\textwidth]{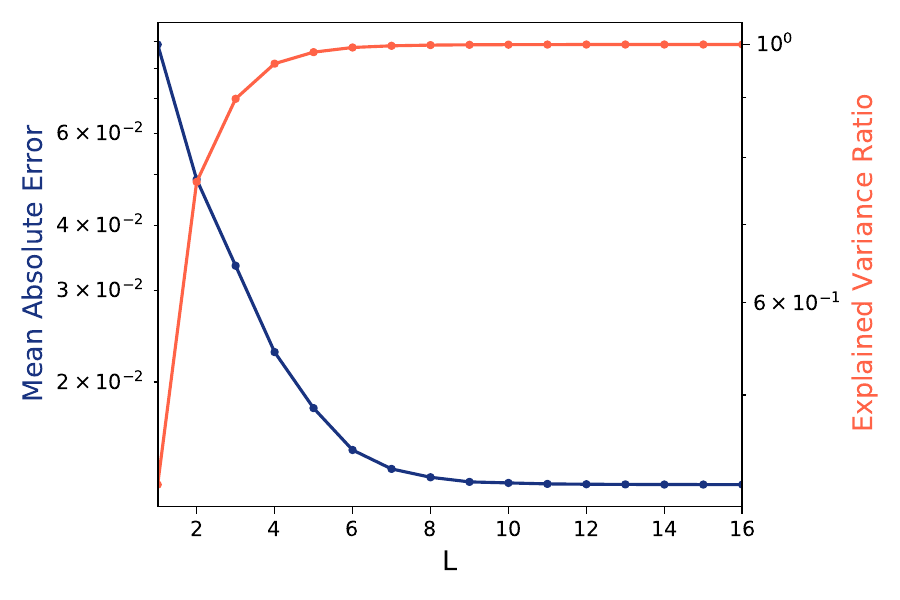}
		\caption{The MAE and the (PCA) explained variance ratio for varying the number of retained functional components $L$ for $K=16$.}
		\label{fig:MAE_in_terms_ofb}
	\end{subfigure}

	\caption[Accuracy of surrogate model in terms of number of functional basis functions and functional components.]{The accuracy (in terms of Mean Absolute Error (MAE) ) of the trained surrogate model on the test dataset, varying the dimensionality reduction dimensions.}
	\label{fig:MAE_in_terms_of} 
\end{figure}

\begin{figure}
		\begin{subfigure}[b]{.49\textwidth}
		\centering
		\includegraphics[width=\textwidth]{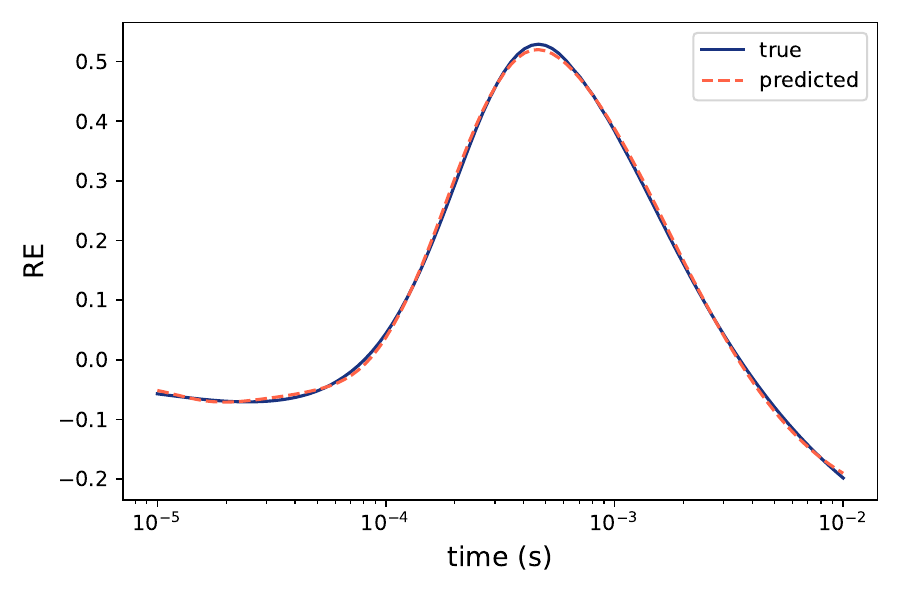}
		\caption{Test case for $d = 50$ m, $\theta = 50.5^\circ$, EC$1$ = 0.47 Sm$^{-1}$, EC$2$ = 0.80 Sm$^{-1}$.}
		\label{fig:casesa}
	\end{subfigure}
	\begin{subfigure}[b]{.49\textwidth}
		\centering
		\includegraphics[width=\textwidth]{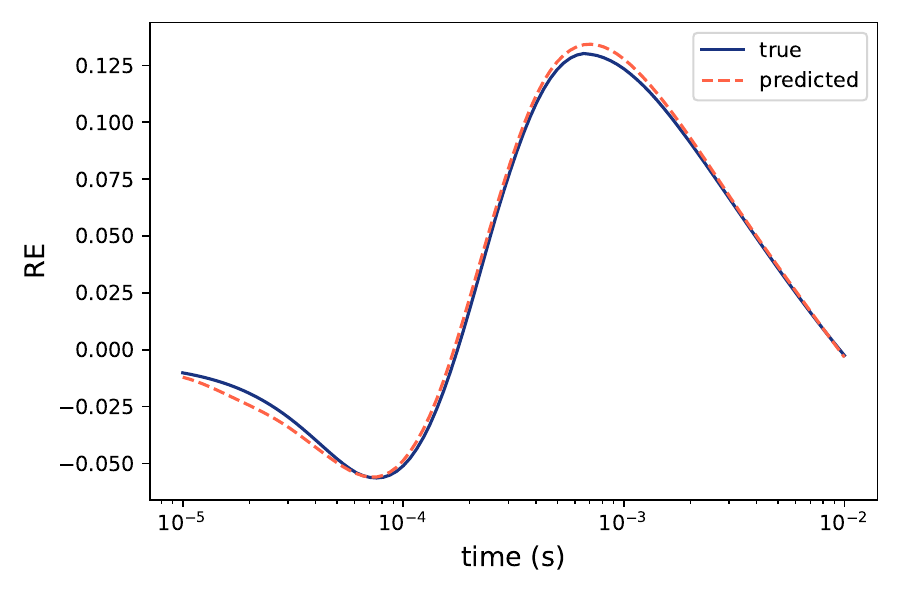}
		\caption{Test case for $d = 29$ m, $\theta = -16^\circ$, EC$1$ = 0.69 Sm$^{-1}$, EC$2$ = 0.89 Sm$^{-1}$.}
		\label{fig:casesb}
	\end{subfigure}
	\begin{subfigure}[b]{.49\textwidth}
		\centering
		\includegraphics[width=\textwidth]{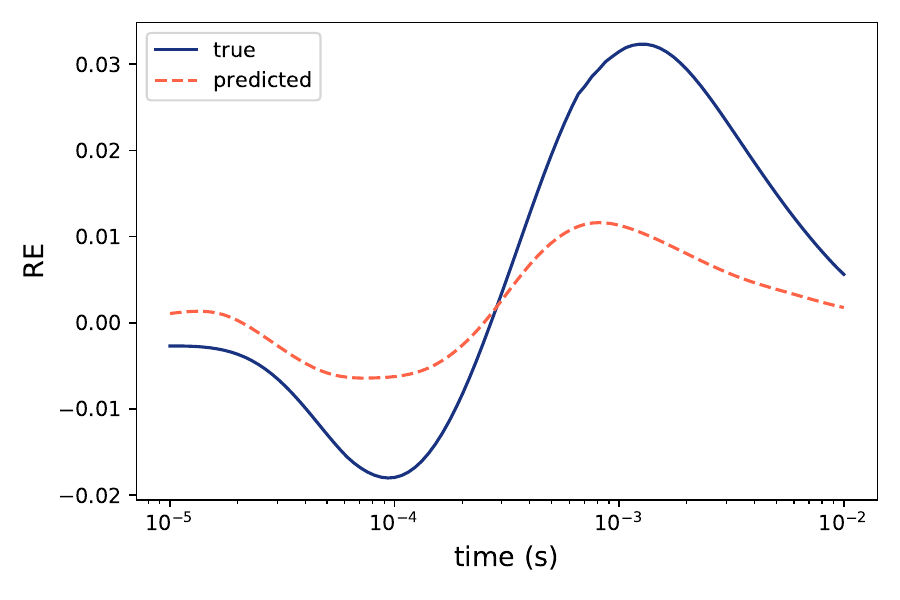}
		\caption{Test case for $d = 11.6$ m, $\theta = -2.33^\circ$, EC$1$ = 0.91 Sm$^{-1}$, EC$2$ = 0.77 Sm$^{-1}$.}
		\label{fig:casesc}
	\end{subfigure}
	\begin{subfigure}[b]{.49\textwidth}
		\centering
		\includegraphics[width=\textwidth]{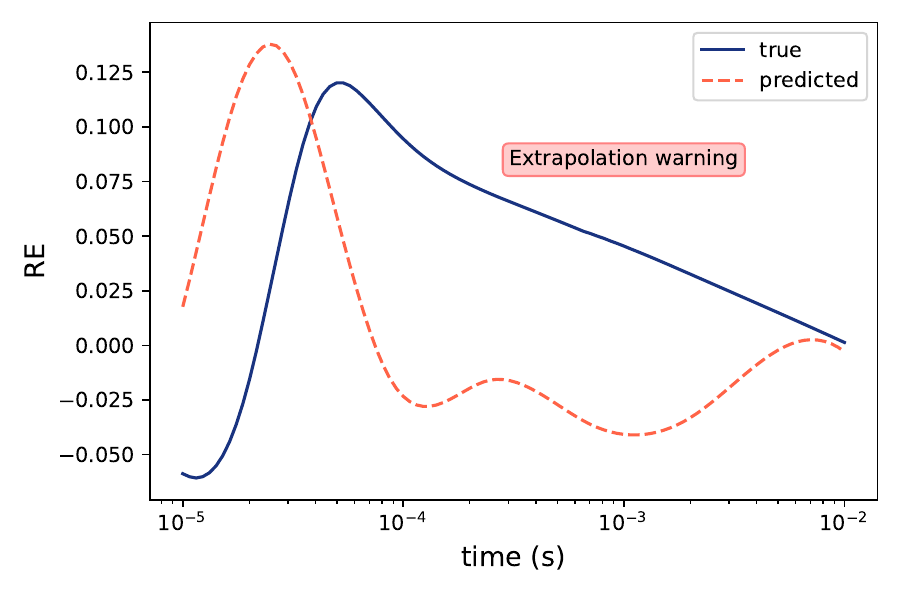}
		\caption{Test case for $d = 37.5$ m, $\theta = 8.3^\circ$, EC$1$ = 0.016 Sm$^{-1}$, EC$2$ = 0.72 Sm$^{-1}$.}
		\label{fig:casesd}
	\end{subfigure}

	\caption{The predicted and the true Relative Error (RE) for four samples from the test dataset (see Table \ref{tab:values}).}
	\label{fig:cases} 
\end{figure}

\begin{table}
	\centering 
	\caption{The features from four samples from the test dataset shown in Figure \ref{fig:cases}.}
	\label{tab:values}
	\begin{tabular}{l|c|c|c|c}
		
		& $d$ (m) & $\theta$ ($^\circ$)  & $EC1$ (S/m)   & $EC2$ (S/m) \\
		\hline
		Case A & 50  & 50.5  &  0.47 &  0.80 \\
		Case B & 29  & -16  & 0.69  & 0.89  \\
		Case C & 11.6 &  -2.33 &0.91  &  0.77 \\
		Case D & 37.5  & 8.3  & 0.016 & 0.72  \\
		
	\end{tabular}
\end{table}

The performance of the trained surrogate model is measured with the MAE in terms of (original) data ($RE$) and the back transform of the reduced data $\tilde{RE}$. The surrogate model is trained for the number of basis functions $K$ in the functional data analysis and the number of functional components $L$ to retain. The parameter $\alpha$ controls the Tikhonov regularization (smoothing) and during optimization, we experienced that a higher value for $\alpha$ was required for functional components that explain less variance (apparently, the White Noise Kernel could not handle the apparent noise properly).  A value of $10^{-5}$ was used for the first four functional components, $10^{-3}$ for the next two and $10^{-2}$ for $l > 6$. 

Based on our insights from the dimensionality reduction in Figure \ref{fig:dimensionality_reduction}, we have optimized the surrogate model for ten different $K$s. The MAE of the predicted data for varying $K$ is shown in Figure \ref{fig:MAE_in_terms_of}A, where the maximum number of functional components $L=K$ is retained. Starting from $K=8$, the prediction accuracy starts to stabilize with $K=16$ as minimum. In Figure \ref{fig:MAE_in_terms_of}B, we show the MAE in terms of number of functional components $L$ for $K=16$. Starting from $L=8$, not much is gained from adding more functional components. Note that this type of optimization leads to a different choice than relying on the (cumulative) explained variance ratio, which is often done for PCA. For the remainder of this text, we proceed with $K=16$ and $L=8$.

To illustrate the reduction in the computational burden, the $HF$ data production requires one hour and 20 minutes of computation time on one 36-core High Performance Computing node (2 x 18-core Intel Xeon Gold 6140 (Skylake @ 2.3 GHz)) and requires 150 GB of RAM. Evaluating the correction model takes two milliseconds on a laptop (MacBook Pro, 10-core M1 Pro chip, 16 GB). The initialization of a pre-trained GPR requires 30 seconds for a training dataset with 5000 samples (this initialization time grows with the number of training samples).

In Figure \ref{fig:cases}, we show some samples illustrating some typical behaviour from the test dataset. Figure \ref{fig:cases}A and B are quite successful, the MAE is 0.00455 and 0.0028 respectively, which is in the order of best MAE given the dimensionality reduction (see Figure \ref{fig:MAE_in_terms_of}). The sample from Figure \ref{fig:cases}C is an example where the angle $\theta$ is small ($-2.33^\circ$). The trend in the predicted relative error $\tilde{RE}$ is good, however the absolute values are too small. The surrogate model thus underestimates the true relative error. The sample in Figure \ref{fig:cases}D is a curious one, as it does not really resemble the typical shape that is observed in the other samples and in the sweeps in Figure \ref{fig:insights}. Yet, it cannot be ascribed to the dimensionality reduction: the MAE of the relative error of the back transformed reduced representation is only 0.00040. The reason can be ascribed to the underlying physics, where the dissipated currents move much more rapidly in more resistive media. This calls for a more thorough sampling of this part of the parameter space.

\begin{figure}
		\begin{subfigure}[t]{.49\textwidth}
		\centering
		\includegraphics[width=\textwidth]{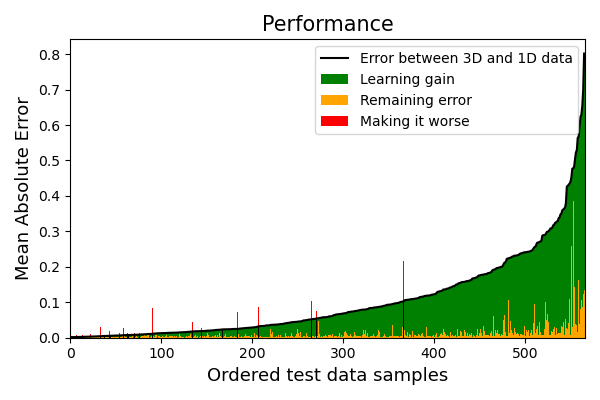}
		\caption{ The learning gain using the surrogate model as correction to the $LF$ model. The black line represents the ordered total MAE of the test samples. The green area shows the total reduced MAE between the 1D ($LF$) and 3D ($HF$) data and the orange area shows the residual MAE of the RE. The cases where the surrogate model makes the prediction of the $HF$ data worse are shown in red.}
		\label{fig:performancea}
	\end{subfigure}
	\begin{subfigure}[t]{.49\textwidth}
		\centering
		\includegraphics[width=\textwidth]{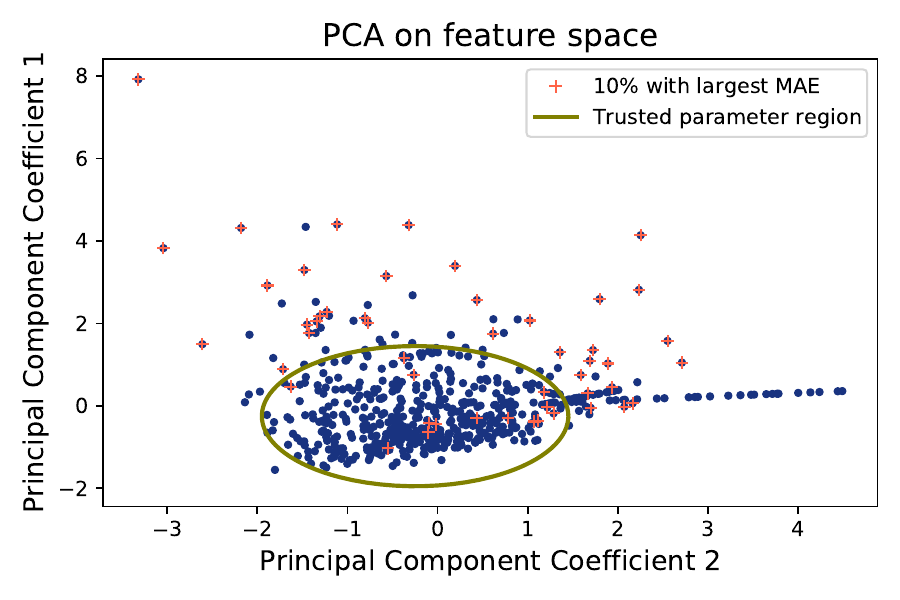}
		\caption{The first and second PC coefficients on the feature space. Samples in the test dataset with the 10\% largest MAE are generally found further from the centre and indicate that the large MAE is a result of being at on the periphery of the trained parameter space. A warning will be printed if the parameters are outside of the trusted region, indicated in green. This is caused by the limited extrapolating capabilities of the surrogate model.  }
		\label{fig:performanceb}
	\end{subfigure}
	\caption[Performance in terms of learning gain and PCA on the feature space.]{Performance in terms of learning gain and PCA on the feature space. }
	\label{fig:performance} 
\end{figure}

\begin{figure}
	\centering
	\includegraphics[width=0.49\textwidth]{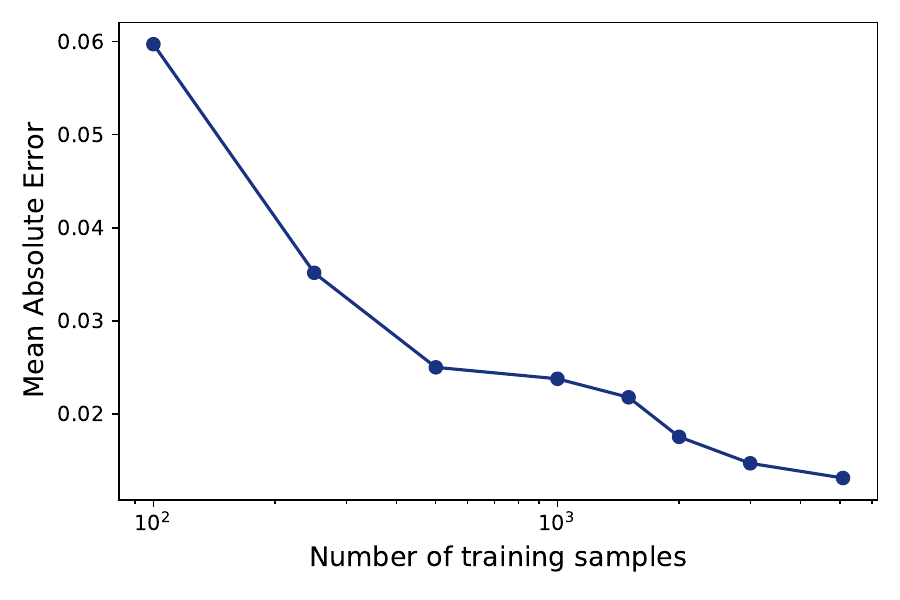}
	\caption[The learning curve.]{The learning curve suggests that more data could increase the overall accuracy.}
	\label{fig:learning_curve} 
\end{figure}
We can also view the performance per test sample separately in Figure \ref{fig:performance}A. The test samples are ordered from small discrepancies between $LF$ and $HF$ data to large discrepancies. The overall green part is what one gains by using the correction/surrogate model to correct the 1D forward $LF$ model. The bottom orange parts indicate the residual error, which remains after applying the surrogate model correction. In some cases, the surrogate model corrects the $LF$ data such that the predicted $HF$ data is worse than when only the $LF$ data was used (red bars), this is in 6\% of the cases. Those cases correspond to low angles, small EC contrasts between layers or a low EC of the first layer and combinations thereof. No particular cause could be identified. In Figure \ref{fig:performance}B we apply PCA to the four input parameters or features. By plotting the first two coefficients, we show the sampled parameter space in reduced dimensions. Samples far from the center (the origin) are extreme values in the sampled data set. The 10\% predictions with the largest MAE are usually in the periphery, suggesting that the limiting extrapolating capabilities of the GPR in combination with the RBF kernel is the cause of the large MAE. Additional sampling in those parameter regions could improve the MAE. A zone of the parameter space was demarcated where the samples of the test data space performed well. We defined the sphere based on the first two PCA components of the feature space, with the median as the centre and a radius as a cut-off parameter. Points outside this sphere lie in a less densely sampled area, and a prediction in that region is less reliable. We have implemented a warning in the accompanying code that alerts when the correction model is used for parameters outside this trusted region. Furthermore, we uncovered that the surrogate correction model performs less satisfactorily at low total MAE, occurring for small angles and for low contrasts in electrical conductivity between the layers. In this case, the maximum relative error at any time is less than 1.5\%, and the surrogate is less relevant. With small angles (less than 2$^\circ$), slight contrasts (less than 0.025) or a combination of these (contrast$^2$ times the angle), the surrogate will warn the end-user that the relative error prediction is less accurate, which is not a problem as the 1D model is then sufficiently accurate. These warnings filter out 62\% of the few test samples for which the MAE after prediction is larger than without using the surrogate model.

Figure \ref{fig:learning_curve} shows the learning curve of the training. It shows the MAE of the GPR accuracy for various training dataset sizes. The MAE was each time computed on the same test dataset of the above correction model. It is of course impossible to make predictions for the course of the curve for more training samples, but it is reasonable to assume that the accuracy can be further improved by providing more training data.

\section{Discussion}
\label{sec:discussion}
As expected, perfect accuracy is not achievable with these straightforward ML techniques, even in this simplified case with two layers. To our knowledge, we lack a measure to determine the overall performance of such a correction model. The effect of a modelling error, e.g., on an inversion model, is still relatively unknown. The approach to look at the learning gain, i.e. to look at what we gain with such correction model compared to only relying on a $LF$ model, requires further validation. However, such a performance measure will be needed. Better accuracy could be achieved by adapting the prior or feature space to the specific application, which will require significant $HF$ data for each project and may not be the most elegant and sustainable approach.

GPR provides relatively simple data fit models, meaning that it interpolates between observations. It is versatile, due to the option for different kernels. The RBF (default) kernel works in most smooth cases. It is difficult to obtain general models, however, and extrapolation is only possible $\ell$ (see Eq. \eqref{eq:RBF}) units away from the trained prior \citep[p.~10]{duvenaud2014automatic}. If one could identify or construct a more closely related underlying kernel, a more general GPR can be constructed with some extrapolating power. This would mean that sampling can be done with a lower data density in the space parameter. Note that GPR's are not sparse and thus use the whole dataset to perform one prediction. If one would have the resources to produce millions of training samples, alternatives to GPR, such as neural networks, should be considered.

In Section \ref{sec:sub:surrogate}, we have explicitly opted to train a single-output GPR on each functional component, rather than using a multiple output GPR. The latter turns out not to be so trivial \citep{wang2015gaussian, liu2018remarks, constantinescu2013physics}. A truly multi-output GPR should not only describe the correlation between data points, but also the correlation between responses \citep{wang2015gaussian}. Hence, it could leverage information from one output to provide a more accurate prediction for another output, rather than modeling them individually \citep{liu2018remarks}. However, multiple-output GPR is still a field of active research and existing codes are sparse. \citet{constantinescu2013physics} point out the difficulty of finding good covariance functions, which turn out to be much more challenging than for single-output GPs.\\

This correction model to the 1D approximation is relevant in contexts with significant modelling errors, such as with large electrical conductivity contrasts between layers and even for relatively small angles ($\theta = 10^\circ$) (see Section \ref{sec:sub:insights}). For example, we are thinking about imaging the freshwater lenses resting on top of salt water, which exhibit large conductivity contrasts and lateral variability. The two-layered forward model may already be sufficient. The correction model can be used with the traditional quasi- or pseudo-2D/3D inversion. The angle is an extra parameter in the inversion and can vary for each sounding. A smoothening constraint could be applied between neighbouring soundings, similar to the height parameter in AEM inversion. 

Alternatively to TDEM applications, the proposed approach could be applied to frequency-domain electromagnetic data. For example, to image the environmental contamination of former gypsum landfills which exhibit large conductivity contrasts with the backfill material \citep{deleersnyder2023estimating} The applicability of the proposed methodology to other geohazard mitigation applications, which tend to become increasingly important in the rising circular economy \citep{koley2022sustainability} and climate change, should be further verified.

\section{Conclusion}
\label{sec:conclusion}
We have presented a novel approach to constructing a surrogate model for 3D forward TDEM simulations with only a few thousand (5090) training samples with relatively simple data-fit models. Rather than predicting the 3D data directly, we predict the relative error between the 3D and 1D data, as the computation of the 1D approximate forward model is fast and explains most of the variability in the geophysical data. By analysing the relative error, we have uncovered that the features of relative error can easily be explained (both height and time of the peak). The contrast in electrical conductivity between the two layers and the deviation from the 1D assumption (via its angle) explain the magnitude of the relative error. In contrast, the interface's depth and the first layer's conductivity determine when the peaks occur. 

The surrogate model in the form of a correction term for the 1D forward model leads to a better prediction of the 3D data. Due to its limited extrapolating capabilities, a warning was implemented when the model is evaluated outside the sufficiently sampled parameter space. In summary, we have successfully tackled the challenge alluded to in the abstract.

Possible improvements could consider (1) active learning, to more knowingly sample the parameter space which maximizes the accuracy gain per extra training sample, especially because the 3D forward simulations are so expensive, (2) generating data on a `Medium fidelity' mesh to increase the number of training samples (while the test dataset should still be generated on an accurate high-fidelity mesh), (3) examine alternative sparse dimensionality reduction techniques (fitting functions to the RE curves did not turn out to perform better, so we did not include these in the current work.). Eventually, as a community, we should define meaningful targets to evaluate the performance of different surrogate models.

\subsection*{Acknowledgments}
The research leading to these results has received funding from FWO (Fund for Scientific Research, Flanders, grant 1113020N and 1113022N) and the King Baudouin Foundation Ernest du Bois prize 2022. The resources and services used in this work were provided by the VSC (Flemish Supercomputer Center), funded by the Research Foundation - Flanders (FWO) and the Flemish Government. We also acknowledge the anonymous reviewer for their comments on this paper.

\subsection*{Code availability section}
The Gaussian Process Regression model is trained with the openly-available Python package scikit-learn. 
The script that controls the above software and produces the correction model are available from \url{https://github.com/WouterDls/correction_term} with the CC-BY 4.0 license.

Contact: wouter.deleersnyder@kuleuven.be

Program language: Python

The datasets are available on Zenodo via \url{https://doi.org/10.5281/zenodo.8366262}.

\bibliographystyle{apalike}
\bibliography{extracted.bib} 

\label{lastpage}

\end{document}